\documentstyle[sprocl,overcite,epsf]{article}

\def\Preprint{\vspace*{-4.8cm} 
  \hfill FTUV/98-16 \\ \mbox{}\hfill 
  IFIC/98-16 \\  \mbox{}\hfill \\ 
  \vspace{3.0cm}}
\def\refjl#1#2#3#4#5#6{\bibitem{#1} #2, {\it #3} {\bf #4} (#5) #6.}

\def\etal{{\it et al}}
\def\NP{Nucl. Phys.}
\def\NPPS{Nucl. Phys. B (Proc. Suppl.)}
\def\PL{Phys. Lett.}
\def\PRL{Phys. Rev. Lett.}
\def\PR{Phys. Rev.}

\def\ZP{Z. Phys.}

\newcommand{\eqn}[1]{(\ref{#1})}
\newcommand{\be}{\begin{equation}}
\newcommand{\ee}{\end{equation}}
\newcommand{\no}{\nonumber}
\newcommand{\bel}[1]{\be\label{#1}}
\newcommand{\ba}{\begin{array}{c}}
\newcommand{\bat}{\begin{array}{cc}}
\newcommand{\ea}{\end{array}}
\newcommand{\beqn}{\begin{eqnarray}}
\newcommand{\eeqn}{\end{eqnarray}}

\newcommand{\bi}{\begin{itemize}}
\newcommand{\ei}{\end{itemize}}

\newcommand{\rms}{\rm\scriptsize}

\newcommand{\cP}{{\cal P}}

\newcommand{\cH}{{\cal H}}
\newcommand{\cA}{{\cal A}}

\begin{document}
\title{ELECTROWEAK PRECISION PHYSICS
 \footnote{Invited talk at the International Workshop 
          {\it Beyond the Standard Model:
          from Theory to Experiment} (Val\`encia, 13--17 October 1997)}
}

\author{A. PICH}

\address{Departament de F\'{\i}sica Te\`orica, 
         IFIC, Universitat de Val\`encia -- CSIC, \\ 
         Dr. Moliner 50, E--46100 Burjassot, Val\`encia, Spain}

\maketitle
\Preprint
\setcounter{footnote}{0}

\abstracts{
Precision measurements of electroweak observables provide
stringent tests of the Standard Model structure and an
accurate determination of its parameters.
A brief overview of the present experimental status is presented.
A more extensive discussion can be found in Ref.~\protect\citen{formigal:97}.
}


\section{Leptonic Charged--Current Couplings}
\label{sec:cc}

\begin{figure}[bth]
\vfill
\centerline{
\begin{minipage}[t]{.45\linewidth}\centering
{\epsfysize =2.5cm \epsfbox{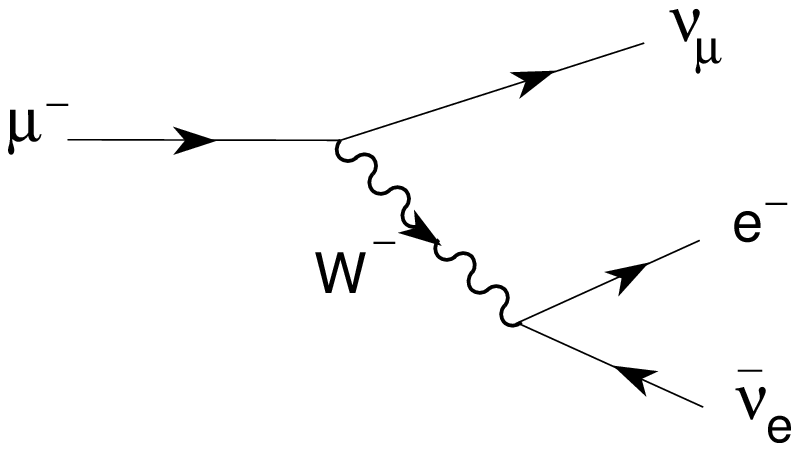}}
\caption{$\mu$--decay diagram.}
\label{fig:mu_decay}
\end{minipage}
\hspace{0.8cm}
\begin{minipage}[t]{.5\linewidth}\centering
{\epsfysize =2.5cm \epsfbox{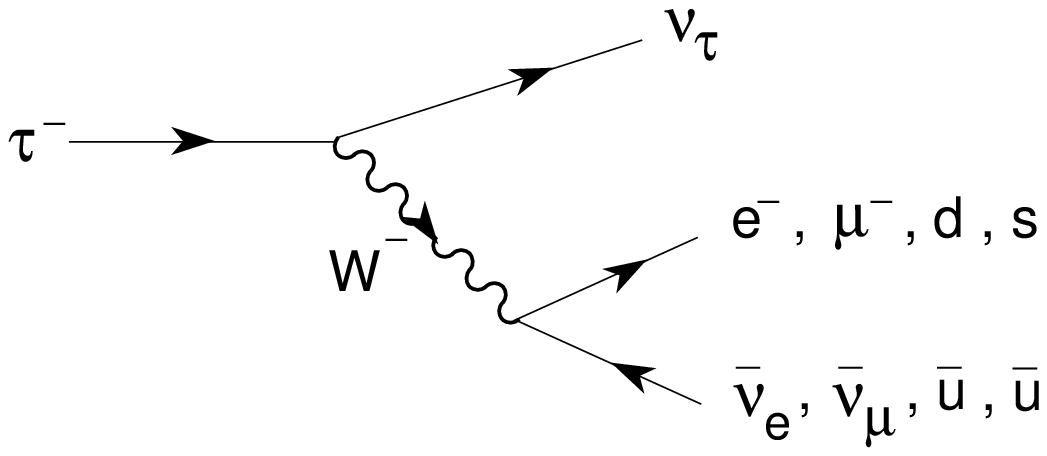}}
\caption{$\tau$--decay diagram.}
\end{minipage}
}
\vfill
\end{figure}

The simplest flavour--changing process is the leptonic
decay of the $\mu$, which proceeds through the $W$--exchange
diagram shown in Figure~\ref{fig:mu_decay}.
The momentum transfer carried by the intermediate $W$ is very small
compared to $M_W$. Therefore, the vector--boson propagator reduces
to a contact interaction.
The decay can then be described through an effective local
4--fermion Hamiltonian,
\bel{eq:mu_v_a}
\cH_{\mbox{\rms eff}}\, = \, {G_F \over\sqrt{2}}
\left[\bar e\gamma^\alpha (1-\gamma_5) \nu_e\right]\,
\left[ \bar\nu_\mu\gamma_\alpha (1-\gamma_5)\mu\right]\, , 
\qquad\quad 
{G_F\over\sqrt{2}} = {g^2\over 8 M_W^2} \, .
\ee
%
The Fermi coupling constant 
$G_F$ 
is fixed by the total decay width,
\bel{eq:mu_lifetime}
{1\over\tau_\mu}\, = \, \Gamma(\mu^-\to e^-\bar\nu_e\nu_\mu)
\, = \, {G_F^2 m_\mu^5\over 192 \pi^3}\,
\left( 1 + \delta_{\mbox{\rms RC}}\right) \, 
f\left(m_e^2/m_\mu^2\right) \, ,
\ee
where
$\, f(x) = 1-8x+8x^3-x^4-12x^2\ln{x}$,
and
$\delta_{\mbox{\rms RC}} = - 0.0042$ 
takes into account the leading higher--order corrections \cite{KS:59,MS:88}.
The measured $\mu$ lifetime \cite{PDG:96},
$\tau_\mu=(2.19703\pm 0.00004)\times 10^{-6}$ s,
implies the value
\bel{eq:gf}
G_F\, = \, (1.16639\pm 0.00002)\times 10^{-5} \:\mbox{\rm GeV}^{-2}
\,\approx\, (293 \:\mbox{\rm GeV})^{-2} \, .
\ee

The leptonic $\tau$ decay widths 
$\tau^-\to e^-\bar\nu_e\nu_\tau,\mu^-\bar\nu_\mu\nu_\tau$
are also given by Eq.~\eqn{eq:mu_lifetime},
making the appropriate changes for the masses of the initial and final
leptons.
Using the value of $G_F$  
measured in $\mu$ decay, one gets a relation between the $\tau$ lifetime
and the leptonic branching ratios \cite{tau96}:
\be\label{eq:relation}
B_{\tau\to e} =  {B_{\tau\to \mu} \over 0.972564\pm 0.000010} = 
{ \tau_{\tau} \over (1.6321 \pm 0.0014) \times 10^{-12}\, \mbox{\rm s} }
\, .
\ee
The errors reflect the present uncertainty of $0.3$ MeV
in the value of $m_\tau$.

\begin{figure}[bt] 
\centering
\vspace{0.3cm}
\centerline{\epsfxsize =9cm \epsfbox{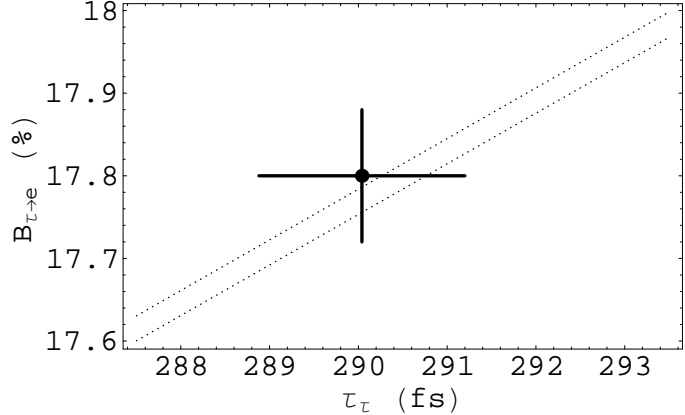}}
\caption{Relation between $B_{\tau\to e}$ and $\tau_\tau$. The dotted
band corresponds to Eq.~(\protect\ref{eq:relation}).
\label{fig:BeLife}}
\end{figure}

The predicted $B_{\tau\to\mu}/B_{\tau\to e}$ 
ratio is in perfect agreement with the measured
value $B_{\tau\to\mu}/B_{\tau\to e} = 0.972 \pm 0.007$.  As shown in
Figure~\ref{fig:BeLife}, the relation between $B_{\tau\to e}$ and
$\tau_\tau$ is also well satisfied by the present data. 
%
These measurements test the universality of the $W$ couplings
to the leptonic charged currents.
Allowing the coupling $g$
to depend on the considered lepton flavour 
(i.e.  $g_e$, $g_\mu$, $g_\tau$), 
the $B_{\tau\to\mu}/B_{\tau\to e}$ ratio 
constrains $|g_\mu/g_e|$, while $B_{\tau\to e}/\tau_\tau$
provides information on $|g_\tau/g_\mu|$.
The present results \cite{formigal:97,tau96}
are shown in Tables \ref{tab:univme} and
\ref{tab:univtm}, together with the values obtained 
from the ratios
$R_{\pi\to e/\mu}\equiv\Gamma(\pi^-\to e^-\bar\nu_e)/
\Gamma(\pi^-\to \mu^-\bar\nu_\mu)$
and 
$R_{\tau/P} \equiv\Gamma(\tau^-\to\nu_\tau P^-) /
 \Gamma(P^-\to \mu^-\bar\nu_\mu)\, $  [$P=\pi,K$],
from the comparison of the $\sigma\cdot B$ partial production
cross-sections for the various $W^-\to l^-\bar\nu_l$ decay
modes at the $p\bar p$ colliders, 
and from the most recent LEP2 measurements of the leptonic
$W^\pm$ branching ratios.  

\begin{table}[tbh]
\centering
\caption{Present constraints on $|g_\mu/g_e|$.}
\label{tab:univme}
\vspace{0.2cm}
\begin{tabular}{|c|c|}
\hline
& $|g_\mu/g_e|$ \\ \hline
$B_{\tau\to\mu}/B_{\tau\to e}$ & $0.9997\pm 0.0037$
\\
$R_{\pi\to e/\mu}$ & $1.0017\pm 0.0015$
\\
$\sigma\cdot B_{W\to\mu/e}$ \ \ ($p\bar p$) & $0.98\pm 0.03$
\\
$B_{W\to\mu/e}$ (LEP2) & $0.92\pm 0.08$
\\ \hline
\end{tabular}\vspace{0.3cm}
%
\caption{Present constraints on $|g_\tau/g_\mu|$.}
\label{tab:univtm}
\vspace{0.2cm}
\begin{tabular}{|c|c|}
\hline
& $|g_\tau/g_\mu|$  \\ \hline
$B_{\tau\to e}\tau_\mu/\tau_\tau$ & $1.0008\pm 0.0030$
\\
$R_{\tau/\pi}$ &  $1.008\pm 0.008$
\\
$R_{\tau/K}$ & $0.997\pm 0.035$
\\
$\sigma\cdot B_{W\to\tau/\mu}$ \ \ ($p\bar p$) & $1.02\pm 0.05$
\\
$B_{W\to\tau/\mu}$ (LEP2) & $1.18\pm 0.11$
\\ \hline
\end{tabular}
\end{table}
%

Although the direct constraints from the measured $W^-\to l^-\bar\nu_l$
branching ratios are meager, the indirect information obtained in
$W^\pm$--mediated decays provides stringent tests of the
$W^\pm$ interactions.
The present data verify the universality of the leptonic
charged--current couplings to the 0.15\% ($\mu/e$) and 0.30\%
($\tau/\mu$) level. The precision of the most recent
$\tau$--decay measurements is becoming competitive with the 
more accurate $\pi$--decay determination. 
It is important to realize the complementarity of the
different universality tests. 
The pure leptonic decay modes probe
the charged--current couplings of a transverse $W$. In contrast,
the decays $\pi/K\to l\bar\nu$ and $\tau\to\nu_\tau\pi/K$ are only
sensitive to the spin--0 piece of the charged current; thus,
they could unveil the presence of possible scalar--exchange
contributions with Yukawa--like couplings proportional to some
power of the charged--lepton mass.

\subsection{Lorentz Structure}
\label{sec:lorentz}

\begin{figure}[tbh] 
\vfill
\centerline{
\begin{minipage}[t]{.47\linewidth}  
\centerline{\epsfxsize =5.64cm \epsfbox{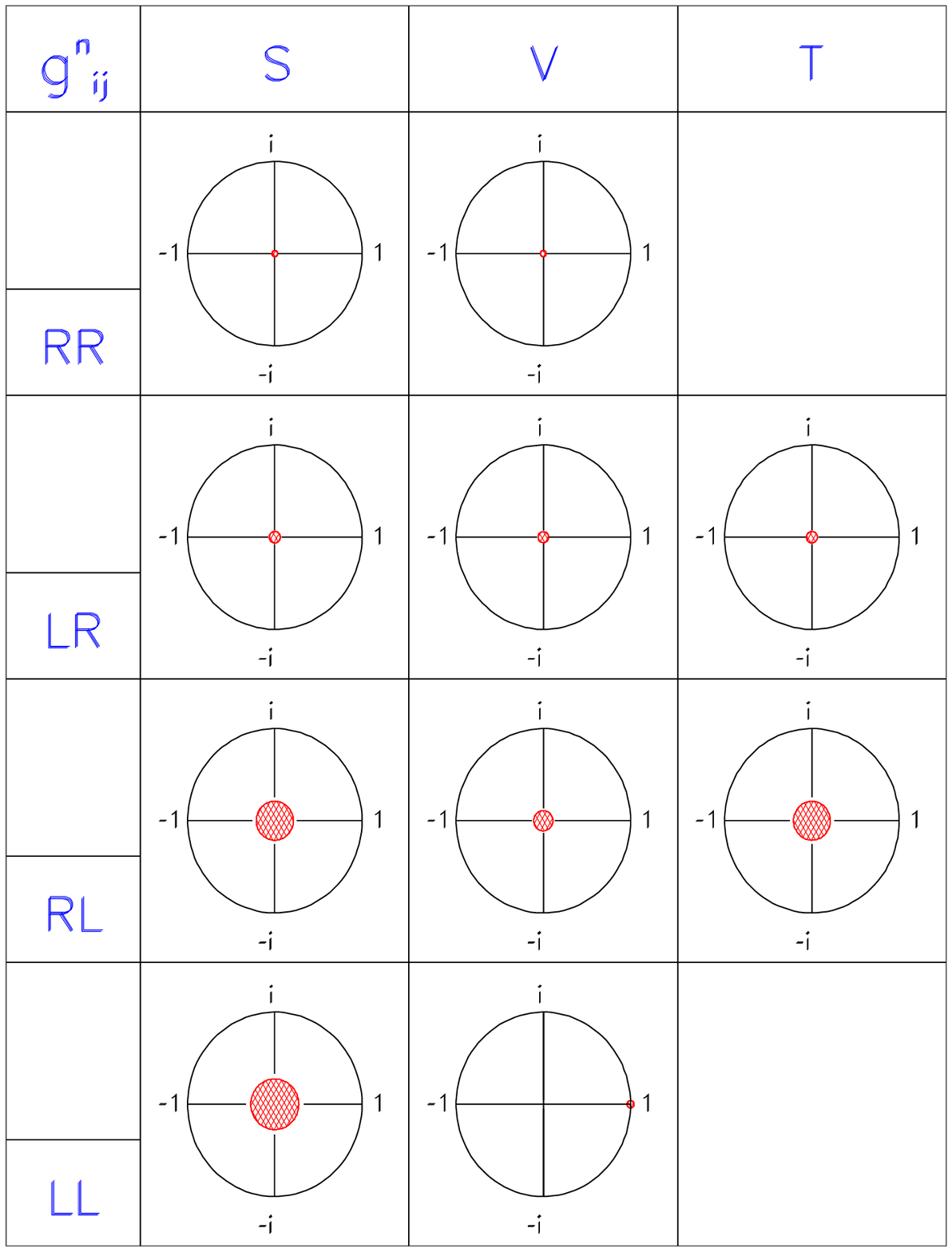}}
\caption{90\% CL experimental limits  \protect\cite{PDG:96}
for the normalized $\mu$--decay couplings
$g'^n_{\epsilon\omega }\equiv g^n_{\epsilon\omega }/ N^n$.
\hfill\hfil }
\label{fig:mu_couplings}
\end{minipage}
\hspace{0.72cm}
\begin{minipage}[t]{.47\linewidth}  
\centerline{\epsfxsize =5.64cm \epsfbox{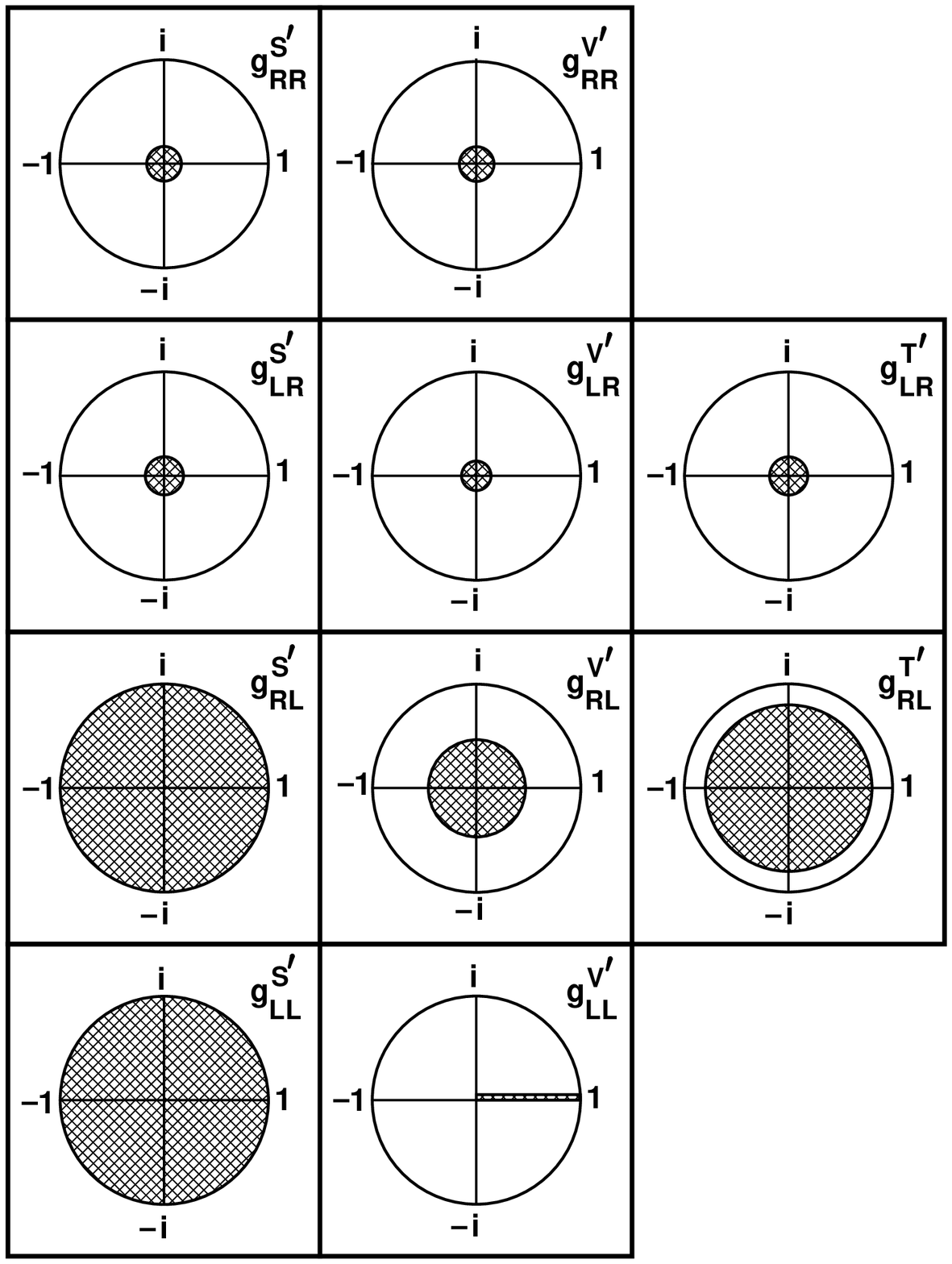}}
\caption{90\% CL experimental limits \protect\cite{cleo:97a}
for the normalized $\tau$--decay couplings
$g'^n_{\epsilon\omega }\equiv g^n_{\epsilon\omega }/ N^n$,
assuming $e/\mu$ universality. \hfill\hfil }
\label{fig:tau_couplings}
\end{minipage}
}
\vfill
\end{figure}

Let us consider the leptonic
decay $l^-\to\nu_l l'^-\bar\nu_{l'}$. 
The most general, local, derivative--free, lepton--number conserving, 
four--lepton interaction Hamiltonian, 
consistent with locality and Lorentz invariance
\cite{MI:50,BM:57,KS:57,FGJ:86,PS:95}
\be
{\cal H} = 4 \frac{G_{l'l}}{\sqrt{2}}
\sum_{n,\epsilon,\omega}          
g^n_{\epsilon\omega}   
\left[ \overline{l'_\epsilon} 
\Gamma^n {(\nu_{l'})}_\sigma \right]\, 
\left[ \overline{({\nu_l})_\lambda} \Gamma_n 
	l_\omega \right]\ ,
\label{eq:hamiltonian}
\ee
contains ten complex coupling constants or, since a common phase is
arbitrary, nineteen independent real parameters.
The subindices
$\epsilon , \omega , \sigma, \lambda$ label the chiralities (left--handed,
right--handed)  of the  corresponding  fermions, and $n$ the
type of interaction: 
scalar ($I$), vector ($\gamma^\mu$), tensor 
($\sigma^{\mu\nu}/\sqrt{2}$).
For given $n, \epsilon ,
\omega $, the neutrino chiralities $\sigma $ and $\lambda$
are uniquely determined.
Taking out a common factor $G_{l'l}$, which is determined by the total
decay rate, the coupling constants $g^n_{\epsilon\omega}$
are normalized to \cite{FGJ:86}
\bel{eq:normalization}
1 = \sum_{n,\epsilon,\omega}\, |g^n_{\epsilon\omega}/N^n|^2 \, ,
\ee
where 
$N^n 
=2$, 1,
$1/\protect\sqrt{3} $ for $n =$ S, V, T.
In the Standard Model (SM), $g^V_{LL}  = 1$  and all other
$g^n_{\epsilon\omega} = 0 $.

The couplings $g^n_{\epsilon\omega}$ can be investigated through the
measurement of the final charged--lepton distribution 
and with the inverse decay
$\nu_{l'} l\to l' \nu_l$. 
For $\mu$ decay, where precise measurements of the polarizations of
both $\mu$ and $e$ have been performed, 
there exist \cite{PDG:96}
stringent bounds on the couplings involving right--handed helicities.
These limits show nicely that 
the $\mu$--decay transition amplitude is indeed of
the predicted V$-$A type:
$|g^V_{LL}|  > 0.96$ (90\% CL).

Figure \ref{fig:tau_couplings} shows the most recent limits
on the $\tau$ couplings \cite{cleo:97a}.
The measurement of the $\tau$ polarization allows to bound those couplings
involving an initial right--handed lepton; however, information on the
final charged--lepton polarization is still lacking. 
The measurement of the inverse decay
$\nu_\tau l\to\tau\nu_l$, needed to separate the $g^S_{LL}$ and
$g^V_{LL}$ couplings, looks far out of reach.

\section{Neutral--Current Couplings}
\label{sec:nc}

In the SM, all fermions with equal electric charge have identical vector, 
$v_f = T_3^f (1-4|Q_f|\sin^2{\theta_W})$ and axial--vector,
$a_f=T_3^f$, couplings to the $Z$ boson.
These neutral current couplings have been precisely
tested at LEP and SLC \cite{lepewwg:97}.

The gauge sector of the SM is fully described in terms of only
four parameters: $g$, $g'$, 
and the two constants characterizing the scalar potential.
We can trade these parameters by \cite{PDG:96,lepewwg:97}
$\alpha$, $G_F$,
\bel{eq:SM_inputs}
M_Z  =  (91.1867\pm 0.0020)\,\mbox{\rm GeV} \, , 
\ee
and $M_H$; this has the advantage of using the 3 
most precise experimental determinations to fix the interaction.
The relations
\bel{eq:A_def}
M_W^2 s_W^2  =  {\pi\alpha\over\sqrt{2} G_F}\, ,
\qquad\qquad
s_W^2  =  1 - {M_W^2\over M_Z^2}\, ,
\ee
determine then $s_W^2 \equiv \sin^2{\theta_W} = 0.2122$ 
and $M_W = 80.94$ GeV;
in reasonable agreement
with the measured $W$ mass \cite{lepewwg:97}, 
$M_W = 80.43\pm 0.08$ GeV.

At tree level, the partial decay widths of the $Z$ boson 
are given by
%
\bel{eq:Z_width}
\Gamma\left[ Z\to \bar f f\right]  =  
{G_F M_Z^3\over 6\pi\sqrt{2}} \, \left(|v_f|^2 + |a_f|^2\right)\, 
N_f  
\, ,
\ee
where $N_l=1$ and $N_q=N_C$.
Summing over all possible final fermion pairs, one predicts
the total width
$\Gamma_Z=2.474$ GeV, to be compared
with the experimental value \cite{lepewwg:97}
$\Gamma_Z=(2.4948\pm 0.0025)$ GeV.
The leptonic decay widths of the $Z$ are predicted to be 
$\Gamma_l\equiv\Gamma(Z\to l^+l^-) = 
84.84$ MeV,
in agreement with the measured value
$\Gamma_l = (83.91\pm 0.10)$ MeV.

Other interesting quantities are 
the ratios
$R_l\equiv\Gamma(Z\to\mbox{\rm hadrons})/\Gamma_l$
and
$R_Q\equiv\Gamma(Z\to\bar Q Q)/ \Gamma(Z\to\mbox{\rm hadrons})$.
The comparison between the tree--level theoretical predictions 
and the experimental values, shown  in
Table~\ref{tab:results}, is quite good.


Additional information can be obtained from the study of the
fermion--pair production process
$e^+e^-\to\gamma,Z\to\bar f f $.
LEP has provided accurate measurements of
the total cross-section, the forward--backward asymmetry,
the polarization asymmetry and the forward--backward polarization
asymmetry:
%
\beqn
\sigma^{0,f}  \equiv  \sigma(M_Z^2)  = 
 {12 \pi  \over M_Z^2 } \, {\Gamma_e \Gamma_f\over\Gamma_Z^2}\, ,
&& \qquad\;
\cA_{\mbox{\rms FB}}^{0,f}\equiv\cA_{FB}(M_Z^2) = {3 \over 4}
\cP_e \cP_f \, ,
\no\\ \label{eq:A_pol_Z}
\cA_{\mbox{\rms Pol}}^{0,f} \equiv
  \cA_{\mbox{\rms Pol}}(M_Z^2)  = \cP_f \, ,\quad
&&  \qquad
\cA_{\mbox{\rms FB,Pol}}^{0,f} \equiv 
\cA_{\mbox{\rms FB,Pol}}(M_Z^2)  =  {3 \over 4} \cP_e  \, ,\quad\quad
\eeqn
where 
$\Gamma_f$ is the $Z$ partial decay width to the $\bar f f$ final state, and
\bel{eq:P_f}
\cP_f \, \equiv \, { - 2 v_f a_f \over v_f^2 + a_f^2} 
\ee
is the average longitudinal polarization of the fermion $f$.

The measurement of the final polarization asymmetries can (only) be done for 
$f=\tau$, because the spin polarization of the $\tau$'s
is reflected in the distorted distribution of their decay products.
Therefore, $\cP_\tau$ and $\cP_e$ can be determined from a
measurement of the spectrum of the final charged particles in the
decay of one $\tau$, or by studying the correlated distributions
between the final products of both $\tau's$ \cite{ABGPR:92}.

With polarized $e^+e^-$ beams, one can also study the left--right
asymmetry between the cross-sections for initial left-- and right--handed
electrons.
At the $Z$ peak, this asymmetry directly measures 
the average initial lepton polarization, $\cP_e$,
without any need for final particle identification.
SLD has also measured the left--right forward--backward asymmetry
for $b$ and $c$ quarks, which are only sensitive to the
final state couplings:
\bel{eq:A_LR}
\cA_{\mbox{\rms LR}}^0\,\equiv\, \cA_{\mbox{\rms LR}}(M_Z^2)
\, = \,  - \cP_e \,  ,
\qquad
\cA_{\mbox{\rms FB,LR}}^{0,f}\,\equiv\,\cA_{\mbox{\rms FB,LR}}^{f}(M_Z^2) 
\, = \, -{3\over 4} \cP_f \, .
\ee
%

\begin{table}[htb]
\begin{center}
\caption{Comparison between 
SM predictions and
experimental \protect\cite{lepewwg:97} measurements. 
The third column includes
the main QED and QCD corrections.
The experimental value for $s_W^2$ refers to the
effective electroweak mixing angle in the charged--lepton sector.
\hfill 
\label{tab:results}}
\vspace{0.2cm}
\begin{tabular}{|c|c|c|c|c|}
\hline
Parameter & \multicolumn{2}{c|}{Tree--level prediction} & 
SM fit & Experimental  
\\ \cline{2-3} & Naive & Improved & (1--loop) & value \\ \hline  
$M_W$ \, (GeV) & 80.94 & 79.96 & $80.375$ & $80.43\pm 0.08$
\\
$s_W^2$ & 0.2122 & 0.2311 & $0.23152$ & $0.23152\pm 0.00023$
\\
$\Gamma_Z$ \, (GeV) & 2.474 & 2.490 & $2.4966$ & $2.4948\pm 0.0025$
\\
$R_l$ & 20.29 & 20.88 & 20.756 & $20.775\pm 0.027$
\\
$\sigma^0_{\mbox{\rms had}}$ \,\, (nb) & 42.13 & 41.38 & 41.467 &
$41.486\pm 0.053$
\\
$\cA_{\mbox{\rms FB}}^{0,l}$ & 0.0657 & 0.0169 & 0.0162 &
$0.0171\pm 0.0010$
\\
$\cP_l$ & $-0.296$ & $-0.150$ & $-0.1470$ & $-0.1505\pm 0.0023$
\\
$\cA_{\mbox{\rms FB}}^{0,b}$ & 0.210 & 0.105 & 0.1031 & $0.0984\pm 0.0024$
\\
$\cA_{\mbox{\rms FB}}^{0,c}$ & 0.162 & 0.075 & 0.0736 & $0.0741\pm 0.0048$
\\
$\cP_b$ & $-0.947$ & $-0.936$ & $-0.935$ & $-0.900\pm 0.050$
\\
$\cP_c$ & $-0.731$ & $-0.669$ & $-0.668$ & $-0.650\pm 0.058$
\\
$R_b$ & 0.219 & 0.220 & 0.2158 & $0.2170\pm 0.0009$
\\
$R_c$ & 0.172 & 0.170 & 0.1723 & $0.1734\pm 0.0048$
\\ \hline
\end{tabular}
\end{center}
\end{table}

Using $s_W^2 = 0.2122$,
one gets the predictions shown in the second column of 
Table~\ref{tab:results}.
The comparison with the experimental measurements
looks reasonable for the total hadronic cross-section 
$\sigma^0_{\mbox{\rms had}} \equiv \sum_q \, \sigma^{0,q}$;
however, all leptonic asymmetries disagree with
the measured values by several standard deviations.
As shown in the table, the same happens with the 
heavy--flavour forward--backward asymmetries
$\cA_{\mbox{\rms FB}}^{0,b/c}$,
which compare very badly with the experimental measurements;
the agreement is however better for $\cP_{b/c}$.

Clearly, the problem with the asymmetries is their high sensitivity
to the input value of $\sin^2{\theta_W}$;
specially the ones involving the leptonic vector coupling
$v_l = (1 - 4 \sin^2{\theta_W})/2$. Therefore, they are an
extremely good window into higher--order electroweak corrections.

\subsection{Important QED and QCD Corrections}
\label{subsec:QED_QCD_corr}

The photon propagator gets vacuum polarization corrections, induced by
virtual fermion--antifermion pairs. Their effect can be
taken into account through a redefinition of the QED coupling,
which depends on the energy scale of the process;
the resulting effective coupling $\alpha(s)$
is called the QED {\it running coupling}.
The fine structure constant is measured
at very low energies; it corresponds to $\alpha(m_e^2)$.
However, at the $Z$ peak, we should rather use $\alpha(M_Z^2)$.
The long running from $m_e$ to $M_Z$ gives rise to a sizeable
correction \cite{ADH:97,EJ:95}:
$\alpha(M_Z^2)^{-1} =   
128.896\pm  0.090\, $.
The quoted uncertainty arises from the light--quark contribution,
which is estimated from $\sigma(e^+e^-\to\mbox{\rm hadrons})$ and 
$\tau$--decay data. 

Since $G_F$ is measured at low energies, while $M_W$ is a
high--energy parameter, the relation between both quantities 
in Eq.~\eqn{eq:A_def}
is clearly modified by vacuum--polarization contributions.
One gets then the corrected predictions
$M_W = 79.96$ GeV and $s^2_W = 0.2311$.

The gluonic corrections to the $Z\to\bar q q$ decays 
can be directly incorporated
by taking an effective number of colours
$N_q = N_C\,\left\{ 1 + {\alpha_s\over\pi} + \ldots\right\}\,
\approx\, 3.12$,
where we have used $\alpha_s(M_Z^2)\approx 0.12\, $.

The third column in Table~\ref{tab:results} shows the numerical impact
of these  QED and QCD corrections. In all cases, the comparison with
the data gets improved. However, it is in the asymmetries where the
effect gets more spectacular. Owing to the high sensitivity to $s^2_W$,
the small change in the value of the weak mixing angle generates
a huge difference of about a factor of 2 in the predicted
asymmetries.
The agreement with the experimental values is now very good.

\subsection{Higher--Order Electroweak Corrections}
\label{subsec:nc-loop}

 Initial-- and final--state photon radiation is by far the
most important numerical correction. One has in addition the contributions
coming from photon exchange between the fermionic lines. 
All these QED corrections are to a large extent dependent on the detector and
the experimental cuts, because of the infra-red problems associated with
massless photons.
These effects are usually estimated with
Monte Carlo programs and subtracted from the data.

More interesting are the so--called {\it oblique} corrections,
gauge--boson self-energies induced by vacuum polarization diagrams,
which are {\it universal} (process independent). 
In the case of the $W^\pm$ and the $Z$, these corrections are sensitive
to heavy particles (such as the top) running along the loop \cite{VE:77}.
In QED, the 
vacuum polarization contribution of a heavy fermion pair
is suppressed by inverse powers of the fermion mass.
At low energies ($s<<m_f^2$), 
the information on the heavy fermions is then lost.
This {\it decoupling} of the heavy fields happens in theories
like QED and QCD,
with only vector couplings and an exact gauge symmetry \cite{AC:75}.
The SM involves, however, a broken chiral gauge symmetry. 
%
The $W^\pm$ and $Z$ self-energies induced by a heavy top
generate contributions 
which increase quadratically with the top mass \cite{VE:77}.
The leading $m_t^2$ contribution to the $W^\pm$ propagator 
amounts to a $-3\% $  correction to the relation \eqn{eq:A_def}
between $G_F$ and $M_W$.

Owing to an accidental $SU(2)_C$ symmetry of the scalar sector,
the virtual production of Higgs particles does not generate any
$m_H^2$ dependence at one loop \cite{VE:77}.
The dependence on the Higgs mass is only logarithmic.
The numerical size of the correction induced on \eqn{eq:A_def}
is $-0.3\% $ ($+1\% $) for $m_H=60$ (1000) GeV.

The vertex corrections 
are {\it non-universal} and usually smaller than the oblique contributions.
There is one interesting exception, the $Z \bar bb$ vertex, which is sensitive
to the top quark mass \cite{BPS:88}.
The $Z\bar f f$ vertex gets 1--loop corrections where a virtual
$W^\pm$ is exchanged between the two fermionic legs. 
Since, the $W^\pm$ coupling changes the fermion flavour, 
the decays
$Z\to \bar d_i\bar d_i$   
get contributions with a top quark
in the internal fermionic lines.
These amplitudes are suppressed by a small quark--mixing factor 
$|V_{td_i}|^2$,
except for the $Z\to\bar b b$ vertex because $|V_{tb}|\approx 1$.
The explicit calculation \cite{BPS:88,ABR:86,BH:88,LS:90}
shows the presence of hard $m_t^2$ corrections to
the $Z\to\bar b b$ vertex,
which amount to a  $-1.5\% $ effect in $\Gamma(Z\to\bar b b)$.

%

The {\it non-decoupling} present in the
$Z\bar b b$ vertex is quite different from the one happening in
the boson self-energies. 
The vertex correction does not have any dependence with the
Higgs mass. Moreover,
while any kind of new heavy particle,
coupling to the gauge bosons, would contribute to the $W^\pm$ and $Z$
self-energies, possible new--physics contributions to the
$Z\bar b b$ vertex are much more restricted and, in any case,
different.
Therefore, an independent experimental test of the two effects
is very valuable in order to disentangle possible
new--physics contributions from the SM corrections.

The remaining quantum corrections (box diagrams, Higgs exchange)
are rather small at the $Z$ peak.


\subsection{Lepton Universality}

\begin{table}[bht]
\centering
\caption{Measured values \protect\cite{lepewwg:97}
of $\Gamma_l$  
and the leptonic forward--backward asymmetries.
The last column shows the combined result 
(for a massless lepton) assuming lepton universality. \hfill
\label{tab:LEP_asym}}
\vspace{0.2cm}
\begin{tabular}{|c|ccc|c|}
\hline
& $e$ & $\mu$ & $\tau$ & $l$ 
\\ \hline
$\Gamma_l$ \, (MeV) & $83.94\pm 0.14$
& $83.84\pm 0.20$ & $83.68\pm 0.24$ & $83.91\pm 0.10$
\\
$\cA_{\mbox{\rms FB}}^{0,l}$ \, (\%) & $1.60\pm 0.24$
& $1.63\pm 0.14$ & $1.92\pm 0.18$ & $1.71\pm 0.10$
\\ \hline
\end{tabular}\vspace{0.2cm}
\centering
\caption{Measured values \protect\cite{lepewwg:97}
of the leptonic polarization asymmetries.}
\label{tab:pol_asym}
\vspace{0.2cm}
\begin{tabular}{|c|c|c|c|}
\hline
$-\cA_{\mbox{\rms Pol}}^{0,\tau} = -\cP_\tau$ &
$-{4\over 3}\cA^{0,\tau}_{\mbox{\rms FB,Pol}} = -\cP_e$ &
$\cA_{\mbox{\rms LR}}^0 = -\cP_e$
& $\!\{{4\over 3}\cA_{\mbox{\rms FB}}^{0,l}\}^{1/2} = -P_l\! $
\\ \hline
$0.1411\pm 0.0064$ & $0.1399\pm 0.0073$ & $0.1547\pm 0.0032$
& $0.1510\pm 0.0044$
\\ \hline
\end{tabular}
\end{table}

\begin{figure}[bth] 
\centerline{\epsfxsize =7cm \epsfbox{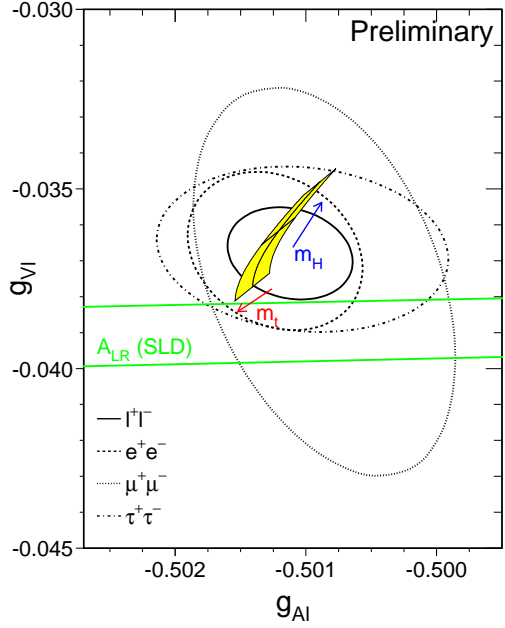}}
\vspace{-0.3cm}
\caption{68\% probability contours in the $a_l$-$v_l$ plane
from LEP measurements \protect\cite{lepewwg:97}. 
The solid contour assumes lepton universality. 
Also shown is the $1\sigma$ band resulting from the
$\protect\cA_{\mbox{\protect\rms LR}}^0$ measurement at SLD. 
The shaded region corresponds to the SM prediction
for \protect{$m_t= 175.6\pm 5.5$} GeV and
\protect{$m_H = 300^{+700}_{-240}$} GeV.
\hfill\hfil }
\label{fig:gagv}
\end{figure}

Tables~\ref{tab:LEP_asym} and \ref{tab:pol_asym}
show the present experimental results
for the leptonic $Z$ decay widths and asymmetries.
The data are in excellent agreement with the SM predictions
and confirm the universality of the leptonic neutral couplings.
There is however a small $1.9\sigma$ discrepancy between the
$\cP_e$ values obtained \cite{lepewwg:97} from 
$\cA^{0,\tau}_{\mbox{\rms FB,Pol}}$ 
and $\cA_{\mbox{\rms LR}}^0$.
The average of the two $\tau$ polarization measurements,
$\cA_{\mbox{\rms Pol}}^{0,\tau}$ and
${4\over 3}\cA^{0,\tau}_{\mbox{\rms FB,Pol}}$,
results in $\cP_l = -0.1406\pm 0.0048$ which disagrees
with the $\cA^0_{LR}$ measurement at the $2.4\sigma$ level.
Assuming lepton universality, 
the combined result from all leptonic asymmetries gives
\bel{eq:average_P_l}
\cP_l = - 0.1505\pm 0.0023 
\qquad (\chi^2/\mbox{\rm d.o.f.}  = 6.0/2) \ .
\ee
Figure~\ref{fig:gagv} shows the
68\% probability contours in the $a_l$--$v_l$ plane,
obtained from  a combined analysis \cite{lepewwg:97} 
of all leptonic observables. 

The neutrino couplings can be determined from the invisible 
$Z$--decay width, 
$\Gamma_{\mbox{\rms inv}}/\Gamma_l = 5.960\pm 0.022$,
by assuming three identical neutrino generations
with left--handed couplings 
and fixing the sign from neutrino scattering 
data \cite{CHARMII:94}.
The resulting experimental value \cite{lepewwg:97},
$v_\nu=a_\nu = 0.50125\pm 0.00092$,
is in perfect agreement with the SM.
Alternatively, one can use the SM prediction, 
$\Gamma_{\mbox{\rms inv}}/\Gamma_l = (1.991\pm 0.001) \, N_\nu$,
to get a determination of the number of (light) neutrino flavours
\cite{lepewwg:97}:
\be
N_\nu = 2.993\pm 0.011 \, .
\ee
The universality of the neutrino couplings has been tested
with $\nu_\mu e$ scattering data, which fixes \cite{CHARMII:94b}
the $\nu_\mu$ coupling to the $Z$: \ 
$v_{\nu_\mu} =  a_{\nu_\mu} = 0.502\pm 0.017$.

Assuming lepton universality,
the measured leptonic asymmetries can be used to obtain the
effective electroweak mixing angle in the charged--lepton sector:
\bel{eq:bar_s_W_l}
\sin^2{\theta^{\mbox{\rms lept}}_{\mbox{\rms eff}}} \equiv
{1\over 4}   \left( 1 - {v_l\over a_l}\right) 
\, = \, 0.23109\pm 0.00029 
\, .
\ee
Including also the information provided by the hadronic
asymmetries, one gets \cite{lepewwg:97} 
$\sin^2{\theta^{\mbox{\rms lept}}_{\mbox{\rms eff}}} = 0.23152\pm 0.00023$
with a $\chi^2/\mbox{d.o.f.} = 12.5/6$.

\subsection{SM Electroweak Fit}

\begin{table}[tbh]
\begin{center}
\caption{Results from the global electroweak fits \protect\cite{lepewwg:97}
to LEP data alone,
to all data except the direct measurements of $m_t$ and $M_W$ at
Tevatron and LEP2, and to all data. \hfill\hfil
\label{tab:EW_fit}}
\vspace{0.2cm}
\begin{tabular}{|c|c|c|c|}
\hline
& LEP only & All data except & All data
\\
& ($M_W$ included) & $m_t$ and $M_W$ &  
\\ \hline 
$m_t$ \,\, (GeV) & $158 {\,}^{+14}_{-11}$ &
$157{\,}^{+10}_{-9}$ & $173.1\pm 5.4$
\\
$m_H$ \,\, (GeV) & $83 {\,}^{+168}_{-49}$ &
$41{\,}^{+64}_{-21}$ & $115{\,}^{+116}_{-66}$
\\
$\log{(m_H)}$ & $1.92 {\,}^{+0.48}_{-0.39}$ &
$1.62{\,}^{+0.41}_{-0.31}$ & $2.06{\,}^{+0.30}_{-0.37}$
\\
$\alpha_s(M_Z^2)$ & $0.121\pm 0.003$ & $0.120\pm 0.003$ & 
$0.120\pm 0.003$
\\ \hline
$\chi^2/\mbox{\rm d.o.f.}$ & $8/9$ & $14/12$ & $17/15$
\\ \hline 

$\sin^2{\theta^{\mbox{\rms lept}}_{\mbox{\rms eff}}}$ &
$0.23188\pm 0.00026$ & $0.23153\pm 0.00023$ & $0.23152\pm 0.00022$ 
\\
$1-M_W^2/M_Z^2$ & $0.2246\pm 0.0008$ & $0.2240\pm 0.0008$ &
$0.2231\pm 0.0006$ 
\\
$M_W$ \,\, (GeV) & $80.298\pm 0.043$ & $80.329\pm 0.041$ &
$80.375\pm 0.030$ 
\\ \hline
\end{tabular}
\end{center}
\end{table}

The high accuracy of the present data provides compelling evidence
for the pure weak quantum corrections, beyond the main QED and QCD corrections
discussed in Section~\ref{subsec:QED_QCD_corr}.
The measurements are sufficiently precise to require the presence of
quantum corrections associated with the 
virtual exchange of top quarks, gauge bosons and Higgses.

Table~\ref{tab:EW_fit} shows the constraints obtained on
$m_t$, $m_H$ and $\alpha_s(M_Z^2)$, from a global fit to the
electroweak data \cite{lepewwg:97}.
The bottom part of the table lists derived results for
$\sin^2{\theta^{\mbox{\rms lept}}_{\mbox{\rms eff}}}$,
$1 - M_W^2/M_Z^2$ and $M_W$.
Three different fits are shown. The first one uses only LEP data,
including the LEP2 determination of $M_W$.
The fitted value of the top mass is in good agreement with the
Tevatron measurement \cite{lepewwg:97}, $m_t= 175.6\pm 5.5$ GeV,
although slightly lower.
The data seems to prefer also a light Higgs.
There is a large correlation (0.76) between the fitted values of $m_t$
and $m_H$; the correlation would be much larger if the $R_b$ measurement
was not used ($R_b$ is insensitive to $m_H$).
The extracted value of the strong coupling agrees very well
with the world average value \cite{PDG:96} $\alpha_s(M_Z^2) = 0.118\pm 0.003$.

The second fit includes all electroweak data except the direct measurements
of $m_t$ and $M_W$, performed at Tevatron and LEP2. The fitted values for these
two masses agree well with the direct determinations. The indirect measurements
clearly prefer low $m_t$ and low $m_H$.
The best constraints on $m_H$ are obtained in the last fit, which includes all
available data. Taking into account additional theoretical uncertainties
due to missing higher--order corrections,
the global fit results in the upper bound \cite{lepewwg:97}:
\bel{eq:M_H}
m_H < 420 \;\mbox{\rm GeV} \qquad (95\% \, \mbox{\rm CL}) \, .
\ee

The uncertainty on $\alpha(M_Z^2)^{-1}$ introduces
a severe limitation on the accuracy of the SM predictions.
To improve the present determination of $\alpha(M_Z^2)^{-1}$
one needs to perform a good measurement of
$\sigma(e^+e^-\to \mbox{\rm hadrons})$, as a function of the centre--of--mass
energy, in the whole kinematical range spanned by DA$\Phi$NE, a 
tau--charm factory and the B factories.
This would result in a much stronger constraint on the Higgs mass.

\vspace{0.5cm}

This work has been supported in part by CICYT (Spain) under grant
No. AEN-96-1718.


\end{document}